\documentclass[aps,prd,twocolumn,superscriptaddress,showpacs,amsmath,amssymb]{revtex4}

\usepackage{graphicx,color}
\usepackage{dcolumn}
\usepackage{subfigure}

\graphicspath{{ps}}

\begin{document}

\title{ \quad\\[1.0cm]\boldmath Study of the suppressed $B$ meson decay 
$B^- \rightarrow DK^-$, $D \rightarrow K^+\pi^-$}

\affiliation{Budker Institute of Nuclear Physics, Novosibirsk}
\affiliation{Chiba University, Chiba}
\affiliation{University of Cincinnati, Cincinnati, Ohio 45221}
\affiliation{The Graduate University for Advanced Studies, Hayama}
\affiliation{Gyeongsang National University, Chinju}
\affiliation{Hanyang University, Seoul}
\affiliation{University of Hawaii, Honolulu, Hawaii 96822}
\affiliation{High Energy Accelerator Research Organization (KEK), Tsukuba}
\affiliation{Hiroshima Institute of Technology, Hiroshima}
\affiliation{Institute of High Energy Physics, Chinese Academy of Sciences, Beijing}
\affiliation{Institute of High Energy Physics, Vienna}
\affiliation{Institute of High Energy Physics, Protvino}
\affiliation{Institute for Theoretical and Experimental Physics, Moscow}
\affiliation{J. Stefan Institute, Ljubljana}
\affiliation{Kanagawa University, Yokohama}
\affiliation{Korea University, Seoul}
\affiliation{Kyungpook National University, Taegu}
\affiliation{\'Ecole Polytechnique F\'ed\'erale de Lausanne (EPFL), Lausanne}
\affiliation{Faculty of Mathematics and Physics, University of Ljubljana, Ljubljana}
\affiliation{University of Maribor, Maribor}
\affiliation{University of Melbourne, School of Physics, Victoria 3010}
\affiliation{Nagoya University, Nagoya}
\affiliation{Nara Women's University, Nara}
\affiliation{National Central University, Chung-li}
\affiliation{National United University, Miao Li}
\affiliation{Department of Physics, National Taiwan University, Taipei}
\affiliation{H. Niewodniczanski Institute of Nuclear Physics, Krakow}
\affiliation{Nippon Dental University, Niigata}
\affiliation{Niigata University, Niigata}
\affiliation{University of Nova Gorica, Nova Gorica}
\affiliation{Osaka City University, Osaka}
\affiliation{Osaka University, Osaka}
\affiliation{Panjab University, Chandigarh}
\affiliation{Saga University, Saga}
\affiliation{University of Science and Technology of China, Hefei}
\affiliation{Seoul National University, Seoul}
\affiliation{Sungkyunkwan University, Suwon}
\affiliation{University of Sydney, Sydney, New South Wales}
\affiliation{Toho University, Funabashi}
\affiliation{Tohoku Gakuin University, Tagajo}
\affiliation{Tohoku University, Sendai}
\affiliation{Department of Physics, University of Tokyo, Tokyo}
\affiliation{Tokyo Institute of Technology, Tokyo}
\affiliation{Tokyo Metropolitan University, Tokyo}
\affiliation{Tokyo University of Agriculture and Technology, Tokyo}
\affiliation{Virginia Polytechnic Institute and State University, Blacksburg, Virginia 24061}
\affiliation{Yonsei University, Seoul}
  \author{Y. Horii}\affiliation{Tohoku University, Sendai}
  \author{K.~Trabelsi}\affiliation{High Energy Accelerator Research Organization (KEK), Tsukuba} 
  \author{H.~Yamamoto}\affiliation{Tohoku University, Sendai} 
  \author{I.~Adachi}\affiliation{High Energy Accelerator Research Organization (KEK), Tsukuba} 
  \author{H.~Aihara}\affiliation{Department of Physics, University of Tokyo, Tokyo} 
  \author{K.~Arinstein}\affiliation{Budker Institute of Nuclear Physics, Novosibirsk} 
  \author{V.~Aulchenko}\affiliation{Budker Institute of Nuclear Physics, Novosibirsk} 
  \author{V.~Balagura}\affiliation{Institute for Theoretical and Experimental Physics, Moscow} 
  \author{E.~Barberio}\affiliation{University of Melbourne, School of Physics, Victoria 3010} 
  \author{I.~Bedny}\affiliation{Budker Institute of Nuclear Physics, Novosibirsk} 
% ...
\collaboration{The Belle Collaboration}

\begin{abstract}
We report a study of the suppressed $B$ meson decay $B^- \rightarrow DK^-$ 
followed by $D\rightarrow K^+\pi^-$, where $D$ indicates a $D^0$ or $\bar{D}^0$ state.
The two decay paths interfere and provide information on the $CP$-violating angle $\phi_3$.
We use a data sample containing $657 \times 10^6~B\bar{B}$ pairs recorded at the 
$\Upsilon(4S)$ resonance with the Belle detector at the KEKB asymmetric-energy $e^+e^-$ 
storage ring.
We do not find significant evidence for the mode $B^- \rightarrow DK^-$, 
$D\rightarrow K^+\pi^-$, and set an upper limit of 
$r_B < 0.19$, where $r_B$ is the magnitude of the ratio of amplitudes 
$|A(B^- \rightarrow \bar{D}^0K^-)/A(B^- \rightarrow D^0K^-)|$. 
The decay $B^- \rightarrow D\pi^-$, $D\rightarrow K^+\pi^-$ is also analyzed as a 
reference, for which we observe a signal with 6.6$\sigma$ significance, 
and measure the charge asymmetry ${\cal A}_{D\pi}$ to be $-0.02^{+0.15}_{-0.16}({\rm stat})\pm 0.04({\rm syst})$.
In addition, the ratio ${\cal B}(B^- \rightarrow D^0K^-)/{\cal B}
(B^- \rightarrow D^0\pi^-)$ is measured to be $[6.77\pm 0.23({\rm stat})\pm 0.30({\rm syst})]\times 10^{-2}$.
\end{abstract}


\maketitle

\tighten

{\renewcommand{\thefootnote}{\fnsymbol{footnote}}}
\setcounter{footnote}{0}

\begin{figure}[htb]
\includegraphics[bb=119 365 464 515, width=0.44\textwidth]{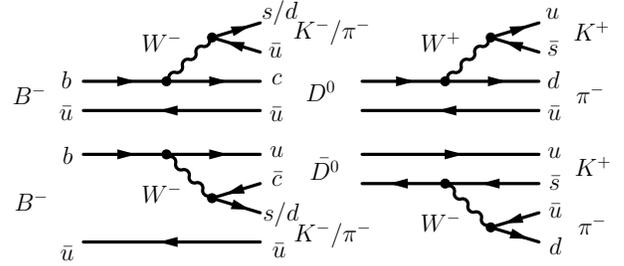}
\caption{ Diagrams for $B^- \rightarrow DK^-$, $D \rightarrow K^+\pi^-$ and 
$B^- \rightarrow D\pi^-$, $D \rightarrow K^+\pi^-$ decays.}
\label{fig:diagram}
\end{figure}

Precise measurements of the parameters of the standard model 
are fundamentally important and may reveal new physics. The 
Cabibbo-Kobayashi-Maskawa matrix~\protect\cite{Cabibbo, KM} consists 
of weak interaction parameters for the quark sector, one of which is 
the $CP$-violating angle $\phi_3 \equiv \arg{(-{V_{ud}}{V_{ub}}^{*}/{V_{cd}}{V_{cb}}^{*})}$. 
Several proposed methods for measuring $\phi_3$ exploit the interference between 
$B^-\rightarrow D^0K^-$ and $B^-\rightarrow \bar{D}^0K^-$, where $D^0$ 
and $\bar{D}^0$ decay to common final states~\protect\cite{DK1,DK2}. 
The effects of $CP$ violation could be enhanced if the final state 
is chosen so that the interfering amplitudes have comparable 
magnitudes~\protect\cite{ADS}. 
The decay $B^- \rightarrow DK^-$, $D \rightarrow K^+\pi^-$ 
($D = D^0~{\rm or}~\bar{D}^0$) is a particularly useful mode, in which
the color-favored $B$ decay followed by the doubly Cabibbo-suppressed $D$ decay 
interferes with the color-suppressed $B$ decay followed by the Cabibbo-favored $D$ 
decay (Fig.~\protect\ref{fig:diagram}). Previous studies of this decay mode have 
not found a significant signal yield~\protect\cite{Saigo,DK_BaBar}.
The decay $B^- \rightarrow D\pi^-$, $D \rightarrow K^+\pi^-$ has a similar event topology and is Cabibbo-enhanced relative to the corresponding $DK^-$ mode.
Therefore this mode is an ideal control sample, while its $CP$ asymmetry is expected to be negligible.

In this analysis, we measure the ratios of the above suppressed decays 
relative to the favored decays $B^- \rightarrow Dh^-$, $D \rightarrow K^-\pi^+$, 
where $h$ $=$ $K$ or $\pi$. The same selection criteria are used for 
the suppressed decays and the favored decays whenever possible in order 
to cancel systematic uncertainties. In this paper, charge conjugate reactions 
are implied except where otherwise mentioned; we denote the suppressed decays 
$B^- \rightarrow Dh^-$, $D \rightarrow K^+\pi^-$ as $B^- \rightarrow D_{\rm sup}h^-$, 
and the favored decays $B^- \rightarrow Dh^-$, $D \rightarrow K^-\pi^+$ 
as $B^- \rightarrow D_{\rm fav}h^-$. Furthermore, a $K^-$ or $\pi^-$ that 
originates directly from a $B^-$ is referred to as the "prompt" particle.

The results are based on a data sample that contains 
$657 \times 10^6~B\bar{B}$ pairs, collected  with the Belle detector 
at the KEKB asymmetric-energy $e^+e^-$ (3.5~GeV on 8~GeV) 
collider~\protect\cite{KEKB} operating at the $\Upsilon(4S)$ resonance.

The Belle detector is a large-solid-angle magnetic
spectrometer that consists of a silicon vertex detector (SVD),
a 50-layer central drift chamber (CDC), an array of
aerogel threshold Cherenkov counters (ACC),  
a barrel-like arrangement of time-of-flight
scintillation counters (TOF), and an electromagnetic calorimeter
comprised of CsI(Tl) crystals (ECL) located inside 
a superconducting solenoid coil that provides a 1.5~T
magnetic field.  An iron flux return located outside of
the coil is instrumented to detect $K_L^0$ mesons and to identify
muons (KLM).  The detector
is described in detail elsewhere~\protect\cite{Belle}.
Two inner detector configurations were used. A 2.0 cm beam pipe
and a 3-layer silicon vertex detector were used for the first sample
of $152 \times 10^6 B\bar{B}$ pairs, while a 1.5 cm beam pipe, a 4-layer
silicon detector, and a small-cell inner drift chamber were used to record  
the remaining $505 \times 10^6 B\bar{B}$ pairs~\protect\cite{svd2}.

Neutral $D$ meson candidates are reconstructed from pairs of oppositely charged tracks.
For each track, we apply a particle identification requirement based on a 
$K/\pi$ likelihood ratio $P(K/\pi) = {\cal L}_K/({\cal L}_K + 
{\cal L}_{\pi})$, where ${\cal L}_K$ and ${\cal L}_{\pi}$ are kaon and pion 
likelihoods, respectively. 
The likelihoods are determined by the information from the ACC and TOF and 
specific ionization measurements from the CDC. We use the requirements 
$P(K/\pi) > 0.4$ and $P(K/\pi) < 0.7$ for the kaon and pion candidates, respectively. 
The efficiency to identify a kaon (pion) is 94\%, while the probability that 
a pion (kaon) is misidentified as a kaon (pion) is about 10\%.
The systematic error in the $K/\pi$ selection efficiency is less than 1\% for both kaons and pions.
The invariant mass of the $K \pi$ pair must be within $\pm 3\sigma$ of the nominal $D$ mass~\protect\cite{PDG}: 1.850 GeV/$c^2 < M(K\pi) <$ 1.880 GeV/$c^2$. 
To improve the momentum determinations, tracks from the $D$ candidate are refitted with 
their invariant mass constrained to the nominal $D$ mass. 

$B$ meson candidates are reconstructed by combining a $D$ candidate with a prompt charged hadron 
candidate, for which the particle identification requirement $P(K/\pi) > 0.6$ 
[$P(K/\pi) < 0.2$] is used for $B^- \rightarrow DK^-$ ($B^- \rightarrow D\pi^-$). 
With this requirement, the efficiency to identify a kaon (pion) is 86\% (81\%), while 
the probability that a pion (kaon) is misidentified as a kaon (pion) is about 5\% (10\%).
The signal is identified by two kinematic variables, the energy difference 
$\Delta E = E_D + E_{h^-} - E_{\rm beam}$ and the beam-energy-constrained mass 
$M_{\rm bc} = \sqrt{\mathstrut E_{\rm beam}^2 - |\vec{p}_D + \vec{p}_{h^-}|^2}$, 
where $E_{\rm beam}$ is the beam energy in the $\Upsilon(4S)$ center-of-mass (c.m.) frame. 
We require $M_{\rm bc}$ to be within $\pm 3\sigma$ of the nominal $B$ mass~\protect\cite{PDG}; 
namely, 5.271 GeV/$c^2 < M_{\rm bc} <$ 5.287 GeV/$c^2$. We then fit the $\Delta E$ 
distribution to extract the signal yield. In the rare cases where there is more 
than one candidate in an event (0.3\% for $B^- \rightarrow D_{\rm sup}K^-$ and 
0.7\% for $B^- \rightarrow D_{\rm sup}\pi^-$), we select the best candidate on 
the basis of a $\chi^2$ determined from the difference between the measured and 
nominal values of $M(K\pi)$ and $M_{\rm bc}$.

The large background from the two jetlike $e^+ e^- \rightarrow q\bar{q}$ 
($q = u,d,s,c$) continuum processes is suppressed using variables that characterize 
the event topology. A Fisher discriminant~\protect\cite{Fisher} made up of modified Fox-Wolfram 
moments called the Super-Fox-Wolfram (SFW)~\protect\cite{KSFW} and $\cos{\theta_B}$, 
where $\theta_B$ is the angle of the $B$ flight direction with respect to the beam 
axis in the c.m. system, are employed. These two independent variables, SFW and 
$\cos{\theta_B}$, are combined to form likelihoods for signal (${\cal L}_{\rm sig}$) 
and for continuum background (${\cal L}_{\rm cont}$); we then construct a likelihood 
ratio ${\cal R} = {\cal L}_{\rm sig}/({\cal L}_{\rm sig} + {\cal L}_{\rm cont})$. 
We optimize the $\cal R$ requirement by maximizing $S/\sqrt{S + B}$, where $S$ and 
$B$ denote the expected numbers of signal and background events in the signal region, 
using Monte Carlo samples.
To estimate $S$, we consider only the contribution from $B^-\rightarrow \bar{D}^0K^-$
followed by $\bar{D}^0 \rightarrow K^+\pi^-$,
where the value of $r_B$ of Eq.~(\protect\ref{eq:rb}) is taken to be 0.1.
For $B^- \rightarrow D_{\rm sup}K^-$ 
($B^- \rightarrow D_{\rm sup}\pi^-$) we require ${\cal R} > 0.90$ (${\cal R} > 0.74$), 
which retains 45\% (70\%) of the signal events and removes 99\% (96\%) of 
the continuum background.
A similar $\cal R$ requirement is obtained if the optimization uses $S/\sqrt{B}$
instead of $S/\sqrt{S + B}$.

For $B^- \rightarrow D_{\rm sup}K^-$, a possible background comes 
from $B^- \rightarrow D\pi^-$, $D\rightarrow K^+K^-$, which has the same final 
state and the same position of the $\Delta E$ peak as the signal.
We veto events that satisfy 1.840 GeV/$c^2 < M(KK) <$ 
1.890 GeV/$c^2$. After this veto, the estimated number of events that 
contribute to the signal yield is $0.22\pm 0.19$.
The favored decay $B^- \rightarrow D_{\rm fav}h^-$ can also produce a 
peaking background for the suppressed decay modes if both the kaon and 
the pion from the $D_{\rm fav}$ decay are misidentified and the particle 
assignments are interchanged. In order to remove this background, we veto events 
for which the invariant mass of the $K\pi$ pair is inside the 1.865 
GeV/$c^2$ $\pm$ 0.020 GeV/$c^2$ window when the mass assignments are 
exchanged. After this requirement, we estimate that $0.17\pm 0.13$ 
($6.0\pm 2.1$) events contribute to the signal yield for 
$B^- \rightarrow D_{\rm sup}K^-$ ($B^- \rightarrow D_{\rm sup}\pi^-$).

The signal yields are extracted using extended unbinned maximum likelihood fits to 
the $\Delta E$ distributions.
For the signal, we use a sum of two Gaussians, where the parameters are determined 
by a fit to $B^- \rightarrow D_{\rm fav}\pi^-$.
The same probability density function (PDF) is used for the signal peaks in all other 
modes; the validity of this assumption is verified by Monte Carlo studies.

Backgrounds from $B\rightarrow XK^-$ ($X\neq D_{\rm sup(fav)}$), such as $B^-\rightarrow D^*K^-$, can populate the negative $\Delta E$ region of the $B^- \rightarrow D_{\rm sup(fav)}K^-$ sample.
The PDF for these backgrounds is obtained from the $B\bar{B}$ Monte Carlo samples, 
in which all known $B$ and $\bar{B}$ meson decays are allowed.
Similarly, backgrounds from $B\rightarrow X\pi^-$ ($X\neq D_{\rm sup(fav)}$), such as $B^- \rightarrow D^*\pi^-$ and $B^- \rightarrow D\rho^-$, can populate the negative $\Delta E$ region of the $B^- \rightarrow D_{\rm sup(fav)}\pi^-$ sample, as well as the negative $\Delta E$ region of the $B^- \rightarrow D_{\rm sup(fav)}K^-$ sample if the prompt pion is misidentified as a kaon.
In the fit to $B^- \rightarrow D_{\rm sup(fav)}\pi^-$ the PDF of these backgrounds is obtained
from the $B\bar{B}$ Monte Carlo samples, while in the fit to $B^- \rightarrow D_{\rm sup(fav)}K^-$
the PDF is obtained from data by assigning the kaon mass to the prompt pion track in the
$B^-\rightarrow D_{\rm sup(fav)}\pi^-$ sample.
The good quality of the fit of the $B^- \rightarrow D_{\rm fav}K^-$ data sample indicates
the validity of this technique.

The feed-across from the $B^- \rightarrow D_{\rm sup(fav)}\pi^-$ signal peak also appears 
in the fit to $B^- \rightarrow D_{\rm sup(fav)}K^-$, where the prompt pion is misidentified 
as the kaon.
The PDF is fixed from the fit to the $B^- \rightarrow D_{\rm fav}\pi^-$ data sample where the kaon mass 
is assigned to the prompt pion track.
The shift caused by the incorrect mass assignment makes the shape of the $\Delta E$ distribution asymmetric, 
and thus we model the misidentification background as a sum of two asymmetric Gaussians, 
for which the left and the right sides have different widths.
In the fit to $B^- \rightarrow D_{\rm sup}K^-$, we fix the yields for the contributions 
from the $B \rightarrow X\pi^-$ background and the feed-across from 
the $B^- \rightarrow D_{\rm sup}\pi^-$ signal peak, using the measured yields in the 
$B^- \rightarrow D_{\rm sup}\pi^-$ sample
scaled by the ratio of the $B^-\rightarrow D_{\rm fav}\pi^-$ yields obtained in the 
$B^-\rightarrow D_{\rm fav}K^-$ and $B^-\rightarrow D_{\rm fav}\pi^-$ analyses.

\begin{figure}[htb]
 \begin{center}
  \leavevmode
  \subfigure
  {\includegraphics[bb=0 0 512 492, width=4.2cm,height=4.2cm]{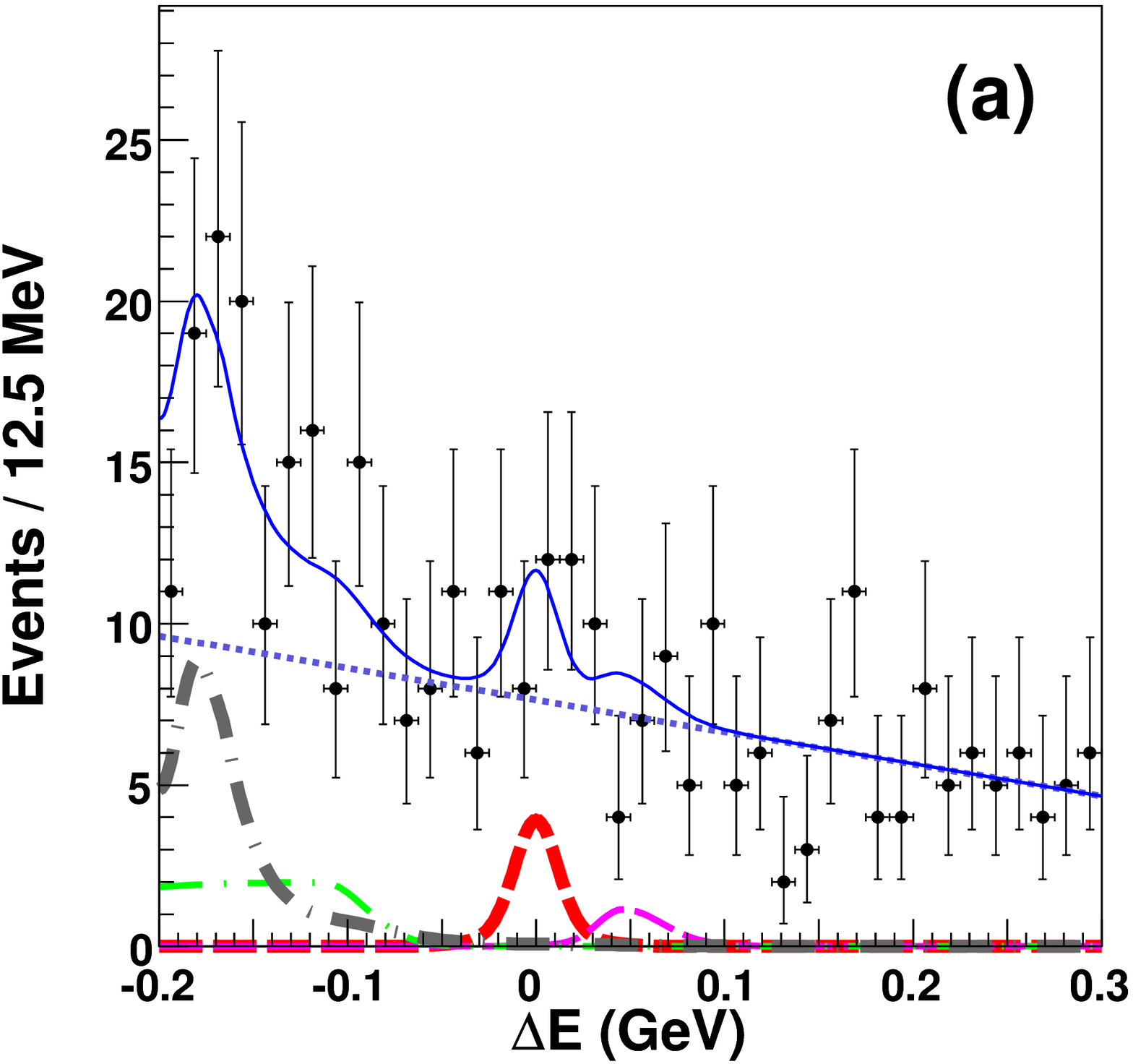}}
  \subfigure
  {\includegraphics[bb=0 0 512 492, width=4.2cm,height=4.2cm]{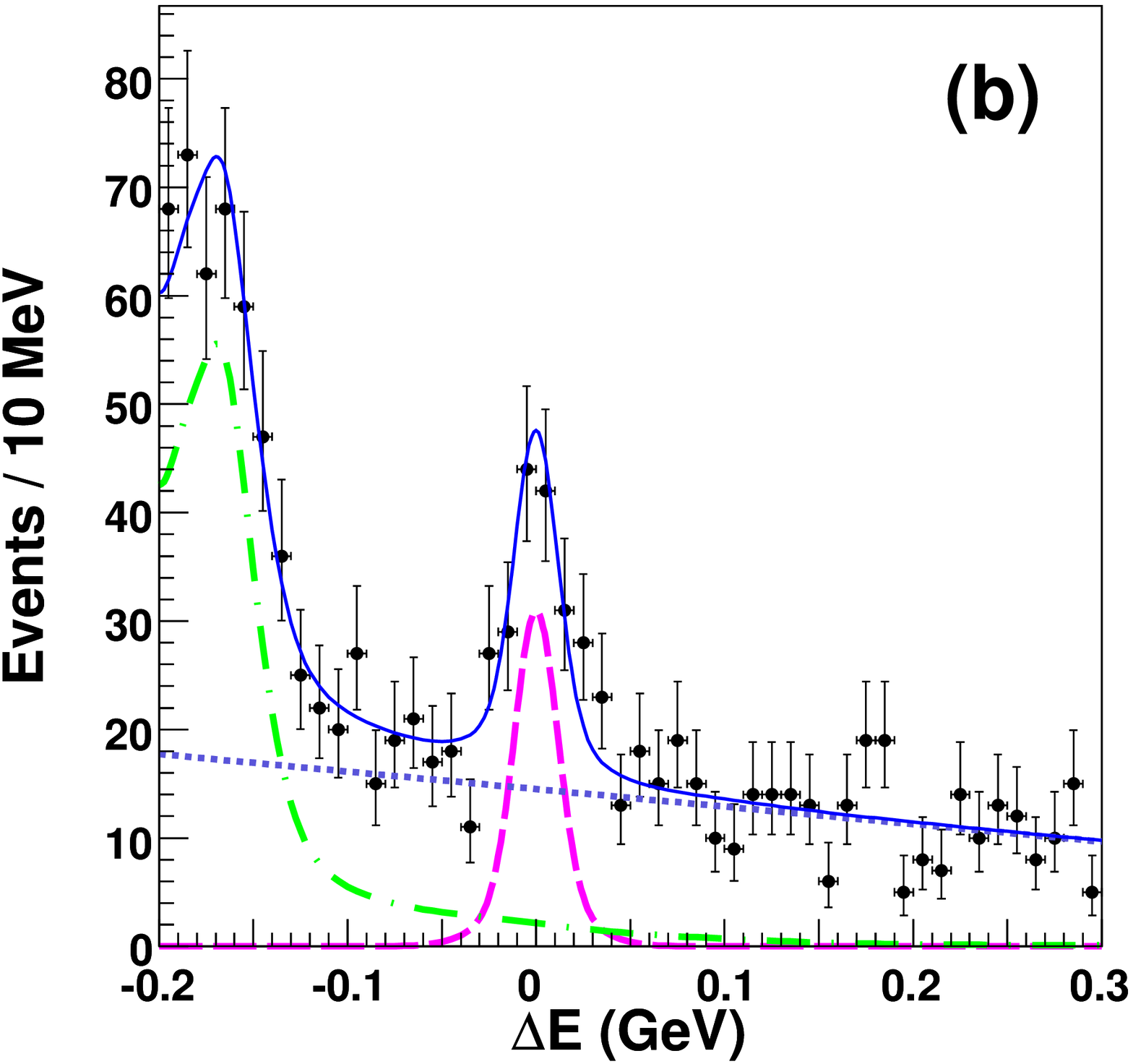}}\\
  \subfigure
  {\includegraphics[bb=0 0 512 492, width=4.2cm,height=4.2cm]{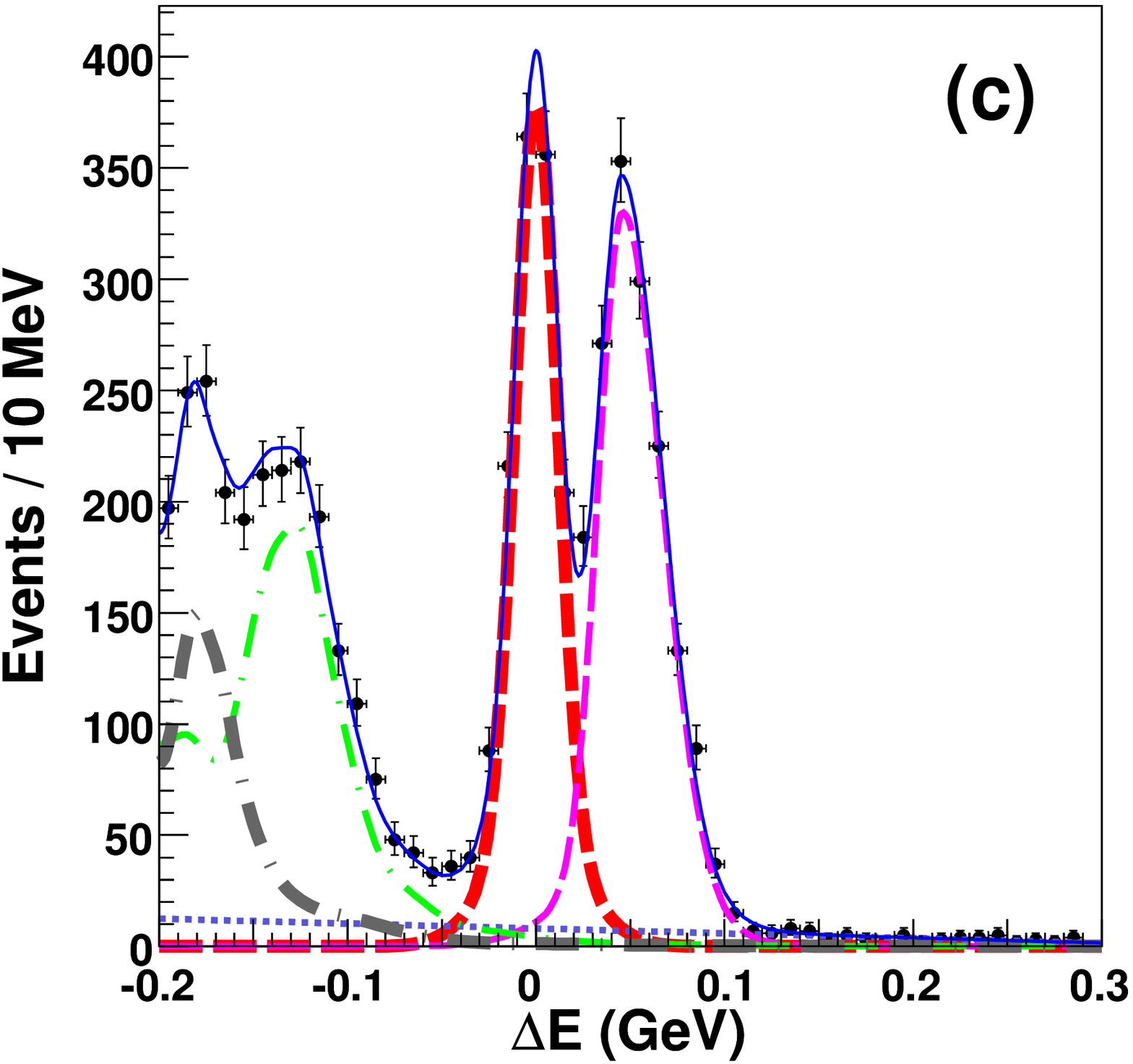}}
  \subfigure
  {\includegraphics[bb=0 0 512 492, width=4.2cm,height=4.2cm]{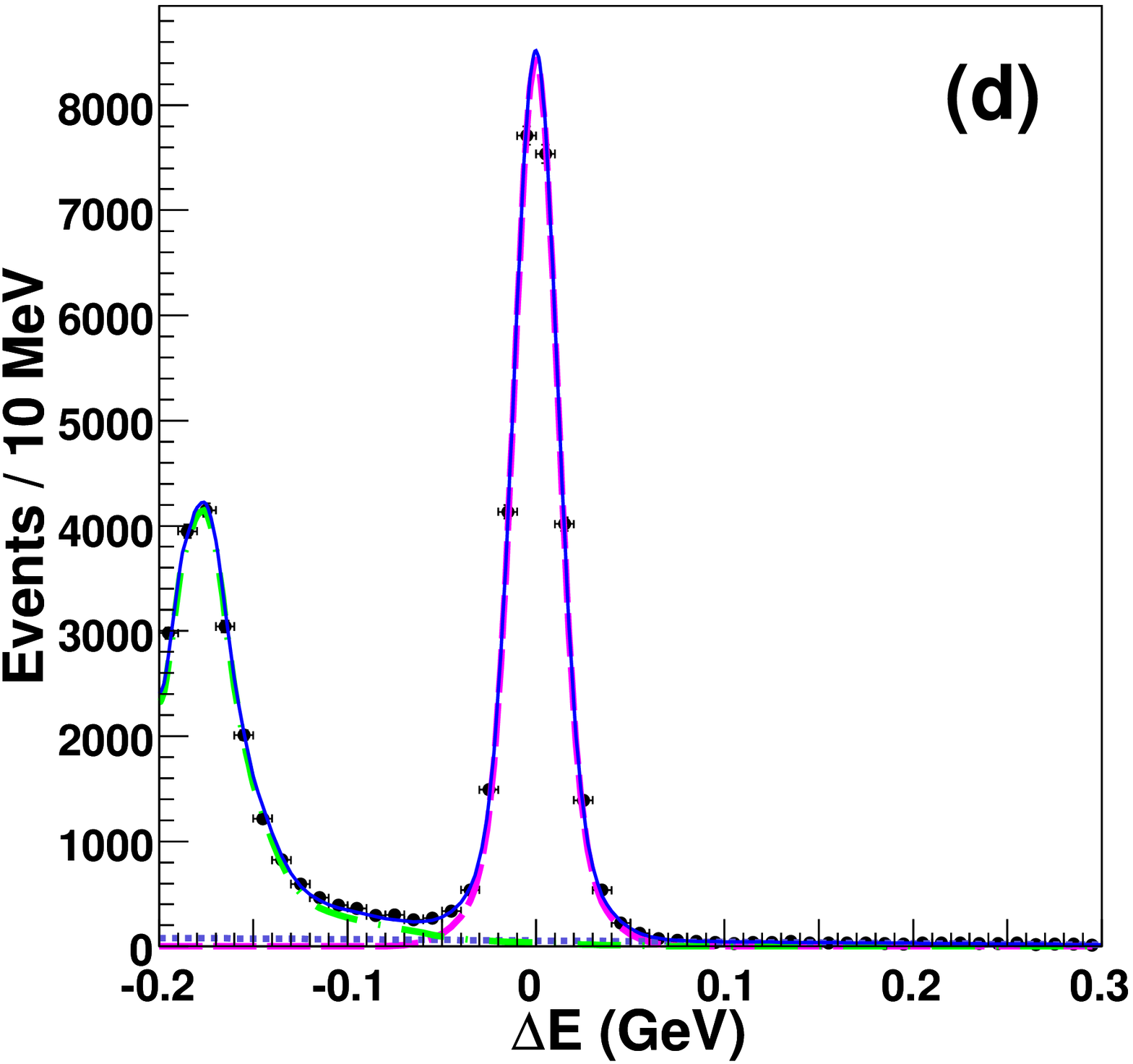}}
  \caption{$\Delta E$ distributions for (a) $B^- \rightarrow D_{\rm sup}K^-$, 
(b) $B^- \rightarrow D_{\rm sup}\pi^-$, (c) $B^- \rightarrow D_{\rm fav}K^-$, and 
(d) $B^- \rightarrow D_{\rm fav}\pi^-$. Charge conjugate decays are included.
In these plots, $B^-\rightarrow DK^-$ components are shown by thicker dashed curves, and $B^-\rightarrow D\pi^-$ components are shown by thinner dashed curves. Backgrounds are shown by thicker dash-dotted curves (for $B\rightarrow XK^-$), thinner dash-dotted curves (for $B\rightarrow X\pi^-$), and dotted curves (for the continuum). The sum of all components is shown by the solid curves.}
  \label{fig:de_fit}
 \end{center}
\end{figure}

The continuum background populates the entire $\Delta E$ region, for which we use a 
linear function.
The fit results are shown in Fig.~\protect\ref{fig:de_fit}.

The charmless decay $B^- \rightarrow K^+K^-\pi^-$ 
($B^- \rightarrow K^+\pi^-\pi^-$) can peak inside the signal region for 
$B^- \rightarrow D_{\rm sup}K^-$ ($B^- \rightarrow D_{\rm sup}\pi^-$).
For this background, we fit the $\Delta E$ distribution of events in the 
$D$ mass sideband, defined as 0.020 GeV/$c^2 < |M(K\pi)-1.865~$GeV/$c^2| <$ 0.080 
GeV/$c^2$, and obtain an expected yield of $-2.3\pm 2.4$ ($2.5\pm 4.5$) events. 
We do not subtract this charmless contribution and instead include the uncertainties, 
$+2.4$ ($+4.5$), in the systematic error.

The signal yields ($N_{Dh^-}$) and the reconstruction efficiencies ($\epsilon_{Dh^-}$) 
for the decays $B^- \rightarrow D_{\rm sup}h^-$ and
$B^- \rightarrow D_{\rm fav}h^-$
are listed in Table~\protect\ref{tb:summary}.
From the results, we calculate ratios of branching fractions, defined as
\begin{eqnarray}
  R_{Dh} \equiv \frac{{\cal B}(B^- \rightarrow D_{\rm sup}h^-)}{{\cal B}(B^- \rightarrow D_{\rm fav}h^-)} = \frac{N_{D_{\rm sup}h^-}/\epsilon_{D_{\rm sup}h^-}}{N_{D_{\rm fav}h^-}/\epsilon_{D_{\rm fav}h^-}}.
\end{eqnarray}
We obtain
\begin{eqnarray}
  R_{DK} &=& [7.8^{+6.2}_{-5.7}({\rm stat})^{+2.0}_{-2.8}({\rm syst})]\times 10^{-3}, \\
  R_{D\pi} &=& [3.40^{+0.55}_{-0.53}({\rm stat})^{+0.15}_{-0.22}({\rm syst})]\times 10^{-3},
\end{eqnarray}
where the systematic errors (Table~\protect\ref{tb:syst}) are subdivided as follows.
\begin{description}
\item[(i)] Fit:
The uncertainties due to the PDFs of the $B^-\rightarrow D_{\rm sup(fav)}h^-$ decays and the $q\bar{q}$ background are obtained by varying the shape parameters by $\pm 1\sigma$.
Those due to the PDFs and yields of the backgrounds from $B\rightarrow XK^-$ and $B\rightarrow X\pi^-$ are estimated by fitting the $\Delta E$ distribution in the region $-0.05$ GeV $< \Delta E <$ $0.15$ GeV without including those contributions.
The total fit error is the quadratic sum and 26\% (3.1\%) for $R_{DK}$ ($R_{D\pi}$).
\item[(ii)] Peaking backgrounds:
The uncertainties due to the backgrounds which peak under the signal were described earlier, and the corresponding systematic error in $R_{DK}$ ($R_{D\pi}$) is estimated to be $^{+~2}_{-25}\%$ ($^{+2.2}_{-5.3}$\%). This uncertainty is asymmetric because the uncertainty of the charmless background is taken only for the negative side.
\item[(iii)] Efficiency:
Monte Carlo statistics and the uncertainties in the efficiencies of particle 
identification requirements dominate the systematic error in detection efficiency, which 
is estimated to be 2.7\% (2.5\%) for $R_{DK}$ ($R_{D\pi}$).
\end{description}
The total systematic error is the sum in quadrature of the above uncertainties.
The possible fit bias is checked using a large number of pseudoexperiments
and found to be negligible.

\begin{table*}[htb]
 \caption{Summary of the fit results. For the $B^- \rightarrow D_{\rm sup}h^-$ signal 
yield, the contribution of peaking backgrounds has been subtracted. The first two errors 
on the measured branching fractions are statistical and systematic, respectively, and 
the third is due to the uncertainty in the $B^- \rightarrow D_{\rm fav}h^-$ branching 
fraction used for normalization. The last column shows the partial rate asymmetries ${\cal A}_{Dh}$ as explained in the text.}
 \label{tb:summary}
 \begin{center}
  \begin{tabular}{lccccc}
  \hline \hline
  Mode~ & ~Efficiency (\%)~ & ~Signal yield~ & ~Significance~ & ~Branching fraction [90\% C.L. upper limit]~ & ~${\cal A}_{Dh}$ \\ \hline
  $B^- \rightarrow D_{\rm sup}K^-$ & 15.4$\pm$0.3 & 9.7$^{+7.7}_{-7.0}$ & 1.3$\sigma$ & 
$(1.2^{+1.0+0.3}_{-0.9-0.4}\pm 0.1)\times 10^{-7}$ & $-0.1^{+0.8}_{-1.0}\pm 0.4$ \\
  &  &  &  & [$2.8\times 10^{-7}$] & \\
  $B^- \rightarrow D_{\rm sup}\pi^-$ & 23.1$\pm$0.4 & 93.8$^{+15.2}_{-14.6}$ & 6.6$\sigma$ 
& $(6.29^{+1.02+0.28}_{-0.98-0.41}\pm 0.24)\times 10^{-7}$ & $-0.02^{+0.15}_{-0.16}\pm 0.04$ \\
  $B^- \rightarrow D_{\rm fav}K^-$ & 15.1$\pm$0.3 & 1220$^{+41}_{-40}$ & $\cdot\cdot\cdot$ 
& $\cdot\cdot\cdot$ & $\cdot\cdot\cdot$ \\
  $B^- \rightarrow D_{\rm fav}\pi^-$ & 22.8$\pm$0.4 & 27202$^{+177}_{-176}$ 
& $\cdot\cdot\cdot$ & $\cdot\cdot\cdot$ & $\cdot\cdot\cdot$ \\
  \hline \hline
  \end{tabular}
 \end{center}
\end{table*}

\begin{table}[htb]
 \caption{Summary of the systematic uncertainties for $R_{Dh}$ and ${\cal A}_{Dh}$.}
 \label{tb:syst}
 \begin{center}
  \begin{tabular}{lcccc}
  \hline \hline
  Source~ & ~$R_{DK}$~ & ~$R_{D\pi}$~ & ~${\cal A}_{DK}$~ & ~${\cal A}_{D\pi}$~  \\ \hline
  Fit & $\pm26$\% & $\pm3.1$\% & $\pm0.40$ & $\pm0.04$ \\
  Peaking backgrounds & $^{+~2}_{-25}$\% & $^{+2.2}_{-5.3}$\% & $\cdot\cdot\cdot$ & $\cdot\cdot\cdot$ \\
  Efficiency & $\pm2.7$\% & $\pm2.5$\% & $\cdot\cdot\cdot$ & $\cdot\cdot\cdot$ \\
  Detector asymmetry & $\cdot\cdot\cdot$ & $\cdot\cdot\cdot$ & $\pm0.01$ & $\pm0.01$ \\
  \hline \hline
  \end{tabular}
 \end{center}
\end{table}

The significances are estimated as $\sqrt{-2\ln{({\cal L}_0/{\cal L}_{\rm max})}}$, 
where ${\cal L}_{\rm max}$ is the maximum likelihood and ${\cal L}_0$ is the 
likelihood when the signal yield is constrained to be zero. 
The distribution of the likelihood $\cal L$ is obtained by convoluting the likelihood in the $\Delta E$ fit and an asymmetric Gaussian whose widths are the negative and positive systematic errors.
The results are shown in Table~\protect\ref{tb:summary}.

Since the signal for $B^- \rightarrow D_{\rm sup}K^-$ is not 
significant, we set an upper limit at the 90\% confidence level (C.L.), 
$R_{DK} < 1.8 \times 10^{-2}$. 
This limit, $R_{DK}^{\rm limit}$, is calculated according to 
$\int_{0}^{R_{DK}^{\rm limit}} {\cal L}(R_{DK}) dR_{DK} = 
0.9 \times \int_{0}^{\infty} {\cal L}(R_{DK}) dR_{DK}$.

Using the values of $R_{Dh}$ obtained above and the 
$B^- \rightarrow D_{\rm fav}h^-$ branching fractions from Ref.~\protect\cite{PDG}, 
we determine the branching fractions for $B^- \rightarrow D_{\rm sup}h^-$ from
\begin{eqnarray}
 {\cal B}(B^- \rightarrow D_{\rm sup}h^-) = 
 {\cal B}(B^- \rightarrow D_{\rm fav}h^-)\times R_{Dh}.
\end{eqnarray}
The results are summarized in Table~\protect\ref{tb:summary}. 
For the $B^- \rightarrow D_{\rm sup}K^-$ branching fraction, we set an upper limit 
at the 90\% C.L., ${\cal B}(B^- \rightarrow D_{\rm sup}K^-) < 2.8\times 10^{-7}$. 
Our branching fraction for $B^- \rightarrow D_{\rm sup}\pi^-$ is consistent with 
the value expected from measured branching fractions for $B$ and $D$ 
decays~\protect\cite{PDG}.

The ratio $R_{DK}$ is related to $\phi_3$ by
\begin{eqnarray}
 R_{DK} = r_B^2 + r_D^2 + 2r_B r_D \cos{\phi_3}\cos{\delta}
\end{eqnarray}
where~\protect\cite{HFAG}
\begin{eqnarray}
 r_B &\equiv& \biggl|\frac{A(B^- \rightarrow \bar{D}^0K^-)}{A(B^- \rightarrow D^0K^-)}
\biggr| \hspace{2mm} , \hspace{3mm} \delta \equiv \delta_B + \delta_D \hspace{1mm} ,  
\label{eq:rb}\\
 r_D &\equiv& \biggl| \frac{A(D^0 \rightarrow K^+\pi^-)}{A(D^0 \rightarrow K^-\pi^+)} 
\biggl| \hspace{1mm} = \hspace{1mm} 0.0578\pm 0.0008,
\end{eqnarray}
and $\delta_B$ and $\delta_D$ are the strong phase differences between the two $B$ and 
$D$ decay amplitudes, respectively. Using the above result,
we obtain a conservative upper limit on $r_B$ as follows.
For a given $R_{DK}$ and in the relevant parameter ranges,
$r_B$ is the largest when $\cos{\phi_3}\cos{\delta} = -1$ and $r_D$ is maximal.
Thus, we take $\cos{\phi_3}\cos{\delta} = -1$ and a $+2\sigma$ shift in $r_D$,
and obtain $r_B < 0.19$ which corresponds to the 90\% upper limit on $R_{DK}$.

\begin{figure}[htb]
 \begin{center}
  \leavevmode
  \subfigure
  {\includegraphics[bb=0 0 512 492, width=4.2cm,height=4.2cm]{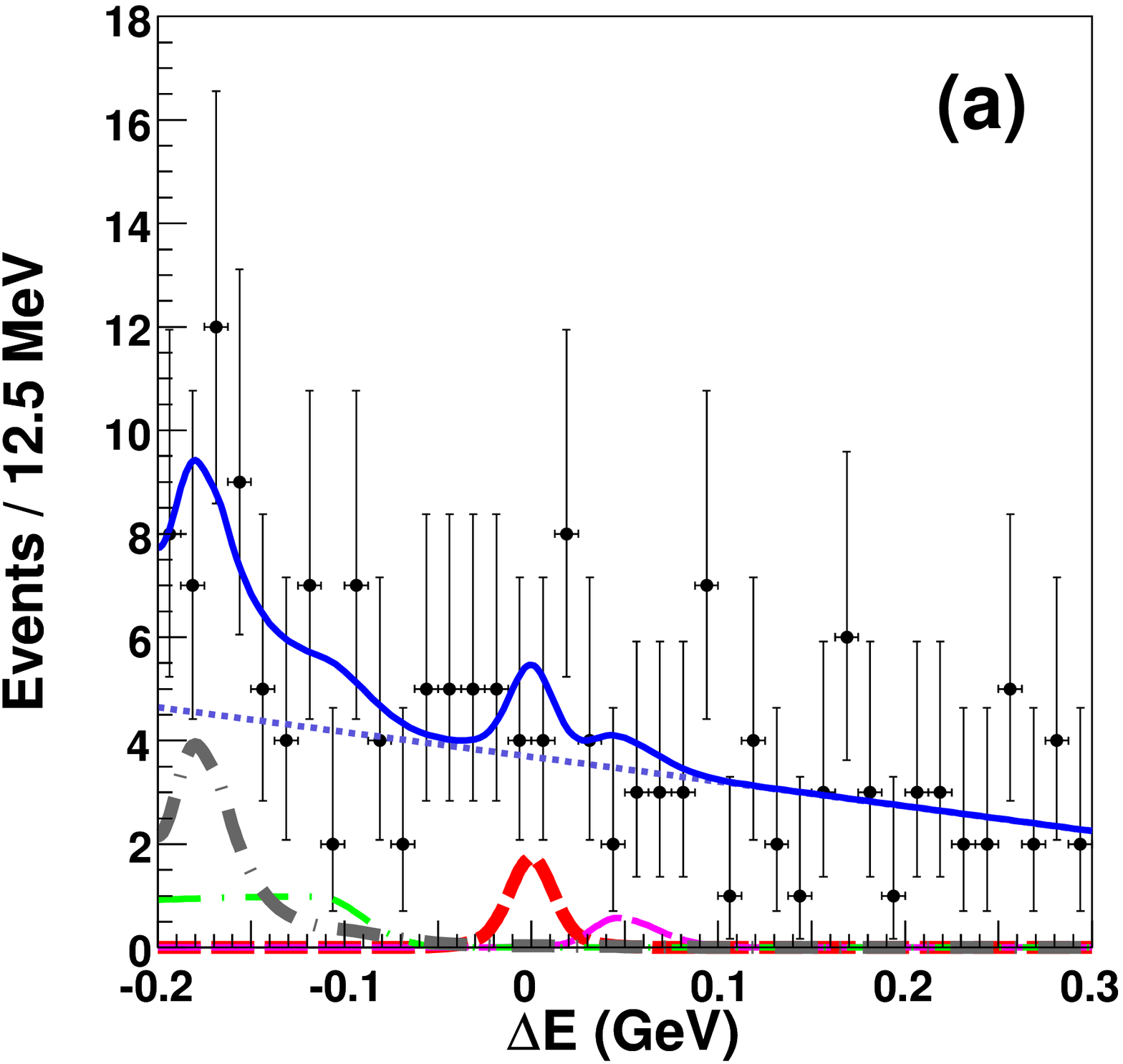}}
  \subfigure
  {\includegraphics[bb=0 0 512 492, width=4.2cm,height=4.2cm]{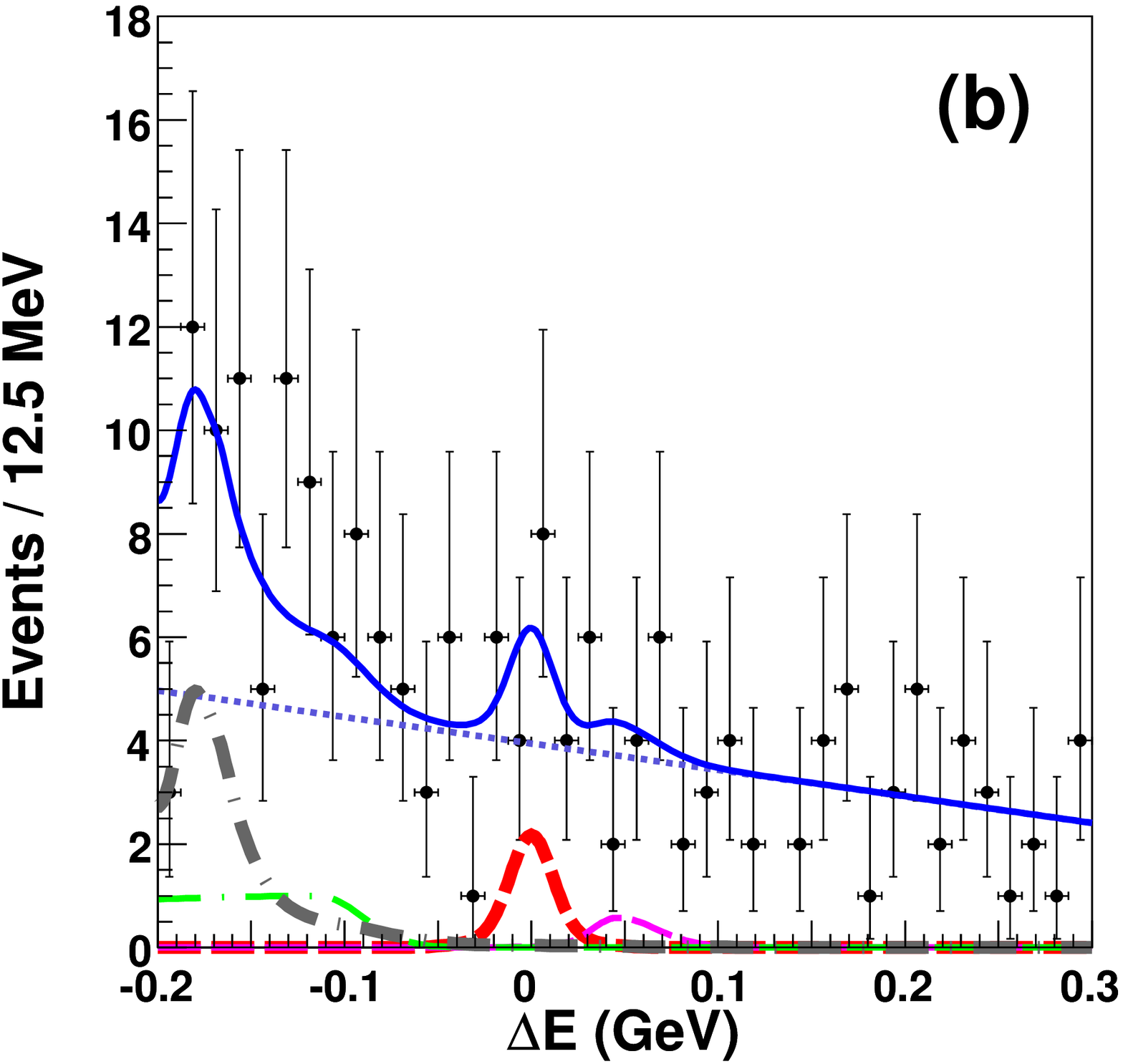}}\\
  \subfigure
  {\includegraphics[bb=0 0 512 492, width=4.2cm,height=4.2cm]{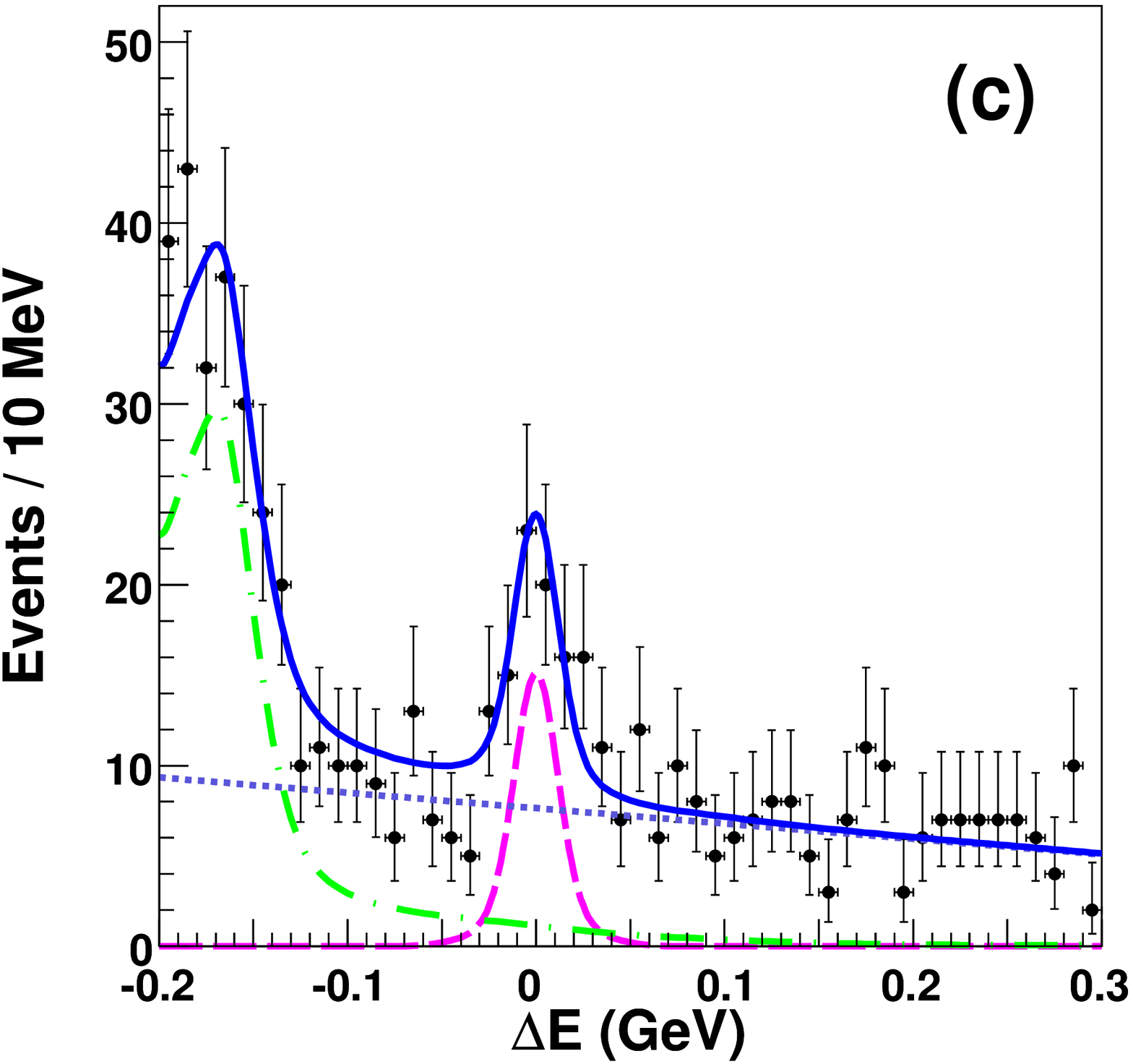}}
  \subfigure
  {\includegraphics[bb=0 0 512 492, width=4.2cm,height=4.2cm]{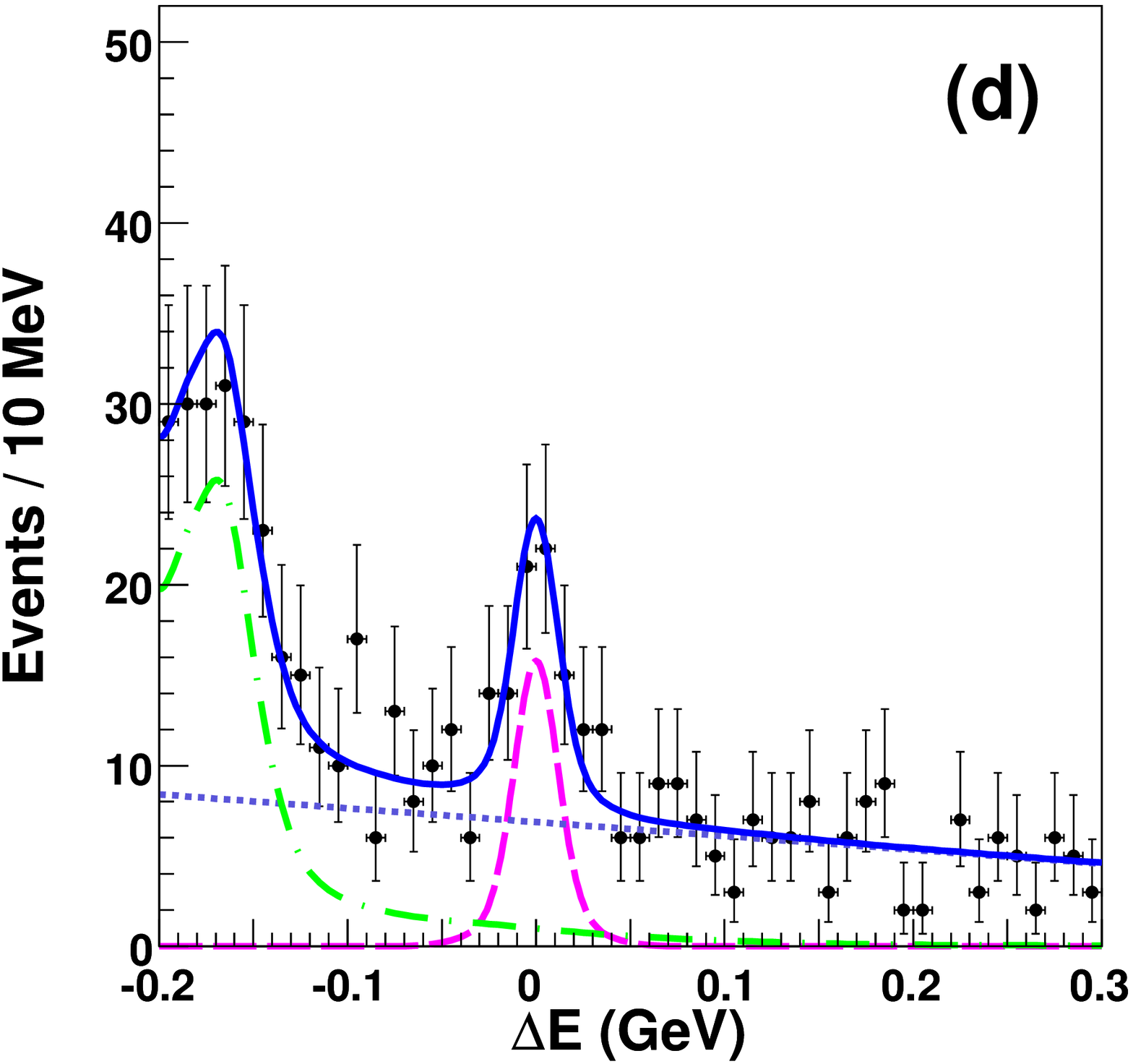}}
  \caption{$\Delta E$ distributions for (a) $B^- \rightarrow D_{\rm sup}K^-$, 
(b) $B^+\rightarrow D_{\rm sup}K^+$, (c) $B^- \rightarrow D_{\rm sup}\pi^-$, and 
(d) $B^+\rightarrow D_{\rm sup}\pi^+$.
The curves show the $B^\mp \rightarrow D_{\rm sup}K^\mp$ component (thicker dashed curves), the $B^\mp \rightarrow D_{\rm sup}\pi^\mp$ component (thinner dashed curves), and the background components (thicker dash-dotted curves for $B\rightarrow XK^\mp$, thinner dash-dotted curves for $B\rightarrow X\pi^\mp$, and dotted curves for the continuum), as well as the overall fit (solid curves).}
  \label{fig:de_fit_asym}
 \end{center}
\end{figure}

We also measure the partial rate asymmetry ${\cal A}_{Dh}$ in 
the $B^{\mp}\rightarrow D_{\rm sup}h^{\mp}$ decays,
\begin{eqnarray}
  {\cal A}_{Dh} &\equiv& \frac{{\cal B}(B^- \rightarrow D_{\rm sup}h^-)-{\cal B}(B^+\rightarrow D_{\rm sup}h^+)}{{\cal B}(B^- \rightarrow D_{\rm sup}h^-)+{\cal B}(B^+\rightarrow D_{\rm sup}h^+)},
\end{eqnarray}
by fitting the $B^-$ and $B^+$ candidates with the asymmetry as one of the fitting parameters.
The fit results are shown in 
Fig.~\protect\ref{fig:de_fit_asym} and included in Table~\protect\ref{tb:summary}. We obtain
\begin{eqnarray}
  {\cal A}_{D\pi} &=& -0.02^{+0.15}_{-0.16}({\rm stat})\pm 0.04({\rm syst})
\end{eqnarray}
and no significant constraint on ${\cal A}_{DK}$.
The systematic errors (Table~\protect\ref{tb:syst}) are dominated
by the uncertainties due to the fits.
Possible bias due to charge asymmetry of the detector is estimated using 
the $B^- \rightarrow D_{\rm fav}\pi^-$ control sample for which the expected asymmetry is small.
The peaking backgrounds are subtracted assuming no $CP$ asymmetries. 
An assumption of 30\% $CP$ asymmetry in the peaking background would lead to 
a shift of 0.02 in ${\cal A}_{D\pi}$.

We also report the ratio
\begin{eqnarray}
  \frac{{\cal B}(B^- \rightarrow D^0K^-)}{{\cal B}(B^- \rightarrow D^0\pi^-)} = \frac{N_{D_{\rm fav}K^-}/\epsilon_{D_{\rm fav}K^-}}{N_{D_{\rm fav}\pi^-}/\epsilon_{D_{\rm fav}\pi^-}}
\end{eqnarray}
to be $[6.77\pm 0.23({\rm stat})\pm 0.30({\rm syst})]\times 10^{-2}$ from the fit 
to $B^- \rightarrow D_{\rm fav}K^-$ and $B^- \rightarrow D_{\rm fav}\pi^-$,
which is about $3\sigma$ lower than the current world average~\protect\cite{PDG}.
The systematic error is due to the uncertainties in the yield extractions (3.1\%) 
and uncertainties in efficiency estimations (1.9\%).
The latter is dominated by the uncertainty in particle identification efficiency for prompt 
hadrons.

In summary, using $657\times 10^{6}$ $B\bar{B}$ pairs collected with the Belle detector, 
we report studies of the suppressed decay $B^- \rightarrow D_{\rm sup}h^-$ ($h = K, \pi$). 
No significant signal is observed for $B^- \rightarrow D_{\rm sup}K^-$ and 
we set a 90\% C.L. upper limit on the ratio of $B$ decay amplitudes, $r_B < 0.19$. 
This result is consistent 
with the measurement of $r_B$ in the Dalitz plot analysis of the decay $B^- \rightarrow DK^-$, 
$D\rightarrow K_S^0\pi^+\pi^-$~\protect\cite{DK_Dalitz,DK_Dalitz_BaBar}. 
For $B^- \rightarrow D_{\rm sup}\pi^-$, we observe a signal with 6.6$\sigma$ significance.
We also report the charge asymmetry for $B^{\mp} \rightarrow D_{\rm sup}\pi^{\mp}$
and the ratio ${\cal B}(B^- \rightarrow D^0K^-)/{\cal B}(B^- \rightarrow D^0\pi^-)$.
These results improve and supersede our previous results~\protect\cite{Saigo, Swain}.

We thank the KEKB group for excellent operation of the
accelerator, the KEK cryogenics group for efficient solenoid
operations, and the KEK computer group and
the NII for valuable computing and Super-SINET network
support.  We acknowledge support from MEXT and JSPS (Japan);
ARC and DEST (Australia); NSFC (China); 
DST (India); MOEHRD, KOSEF and KRF (Korea); 
KBN (Poland); MES and RFAAE (Russia); ARRS (Slovenia); SNSF (Switzerland); 
NSC and MOE (Taiwan); and DOE (USA).

\end{document}